# What makes us a community: structure, correlations, and success in scientific world

Sergei V. Kalinin[1,2] and Artem Maksov[3]

[1] The Institute for Functional Imaging of Materials, Oak Ridge National Laboratory, Oak Ridge, TN 37831

[2] The Center for Nanophase Materials Sciences, Oak Ridge National Laboratory, Oak Ridge, TN 37831

[3] UT/ORNL Bredesen Center, University of Tennessee, Knoxville, TN

We explore the statistical structure of scientific community based on multivariate analysis of publication (or other identifiable metrics) distribution in the author space. Here, we define community based on keywords, i.e. projecting semantic content of the documents on predefined meanings; however, more complex approaches based on semantic clustering of publications are possible. Remarkably, this simple statistical analysis of publication metadata allows understanding of internal interactions with community in general agreement with experience acquired over decades of social interaction within it. We further discuss potential applications of this approach for ranking within the community, reviewer selection, and optimization of community output.

Scientists are humans, and as humans they are collective. We talk about our community and our peers, we care about standing and reputation in the community. Intuitively, these notions seem to be fundamental and obvious. However, any attempt to rigorously describe or quantify it encounters significant problems.[1] For example, given the role the standing in the community plays in individuals' career advancement, multiple attempts have been made to introduce rigorous metrics of scientific productivity, ranging from number of publications to number of citations to *h*-index. However, it is recognized that these metrics cannot be compared directly between different communities due to sufficiently different measures of productivity, accepted publications types, etc and can even negatively impact community.[2] Here "community" would be defined as "biologists" or "condensed matter experimental physicists" or so – sufficiently large groups of scientists having common language, attending similar conferences, and maintaining continuous professional and personal interactions. In the most simplistic term, the community can be defined based on overarching topic or set of keywords (external basis). In more advanced version, it can be defined based on hierarchical clustering of the semantic context of publications[3, 4] or professional social network[5] (unsupervised clustering)[6] that further allow analysis of citation or meme propagation.[7] Here, we will not attempt to define community precisely and rather define interactions and ranking within a community defined via defined keywords.

Here, we propose an approach based on the multivariate statistical analysis of the connectivity in the scientific world. We attempt to build these definitions on unambiguously defined parameters, and attempt to establish relationships with more subjective identifiers that can vary between different fields and cultures.

As a starting point in building this formalism, we need to identify the measure of activity in scientific community. These can be quantified through multiple parameters, including publications, conference attendance, time served at the committees, or even direct interaction. Here, we focus the analysis on the peer reviewed publications, both as a recognized standard and easily traceable through multiple data bases. However, more complex aggregate metrics can be introduced; alternatively, correlations between different metrics can be explored.

We further proceed to identify an individual contributor, i.e. a single author, as a fundamental element of a scientific community. We aim to understand and analyze the contribution of an individual person to the community, and how these contributions actually form the community. Similarly to author lists of the publication, the authors are uniquely defined, e.g. using ORCHID or equivalent systems

Correspondingly, we can define a publication space as an N-dimensional space with each dimension corresponding to an individual author. The dimensionality of the space, N, is then defined by the total number of ever lived and living scientists and is large, but finite. Each publication in this case then defines a single point in this space. For example, for the 28-dimensional space of authors of {A,B,C, ... , Z} the publication authored by A and C will be represented as {1,0,0, ..., 0}, and by A,B, and Z as {1,1,0, ..., 0,1}, where all missing elements are zeroes. We further note that this definition can be easily adapted to account for relative contributions, i.e. equal contribution can correspond to {0.5, 0, 0.5, 0...} and so on. The division of the credit within the paper cannot performed *a priory* based on publication only, and either have to be postulated (e.g. ascribing certain weights depending on position within the author list), or determined based on the internal author dynamics.

We therefore proceed to analyze the statistical properties of the distributions of the publications in the author space. Here, we perform the ISI search for the keyword "piezoresponse force microscopy", defining a well-defined community exploring ferroelectric materials by voltage –modulated scanning probe microscopy. The search yields 955 references (as of 7/21/2014). The statistical analysis of the authors yields frequency distribution referred in Table I. Here, we remove author' specific information to focus on the salient aspects of observed behavior.

**Table I**

Author analysis based on contributions

| Top authors | Contributions | Total eigenvectors | Total contribution | Scaled by variation |
|---|---|---|---|---|
| A | 95 | 470 | 519.3 | 449.8 |
| B | 63 | 457 | 477.6 | 334.9 |
| C | 38.75 | 481 | 483.4 | 274.7 |
| D | 38.25 | 491 | 583.5 | 388.0 |
| E | 28.5 | 477 | 641.8 | 288.6 |
| F | 28.25 | 480 | 632.8 | 233.2 |
| G | 25 | 468 | 533.1 | 202.8 |
| H | 23.5 | 481 | 691.3 | 177.7 |
| I | 22.25 | 471 | 459.8 | 145.2 |
| J | 20.75 | 450 | 683.1 | 207.8 |
| K | 20.25 | 483 | 672.9 | 240.9 |
| L | 19.25 | 456 | 680.7 | 225.9 |
| M | 18.5 | 463 | 650.1 | 191.4 |
| N | 17.75 | 450 | 685.4 | 196.6 |
| O | 17.25 | 497 | 773.4 | 159.6 |
| P | 15 | 457 | 566.4 | 135.5 |
| Q | 15 | 455 | 550.0 | 115.4 |
| R | 14.25 | 466 | 733.0 | 178.2 |
| S | 13.75 | 485 | 632.7 | 133.5 |
| T | 13.5 | 459 | 708.1 | 143.3 |

Table I demonstrates the descriptive metrics of authors in the field described by this keyword. First column contains author names (or rather indices). Second column contains total count of papers as scaled by author position in the publication, where first and last author received a 1.0 score per paper, second and second from last received 0.5, and all else received 0.25 (note that this choice of weighting factors is *ad hoc*, and below we discuss some of possible strategies for selection of weighting schemes). Third column contains number of eigenvectors for a given author, in which their contribution was calculated as positive. Fourth column contains sum of total contribution to all eigenvectors in previous metric. Finally, last column shows sum of total contribution to all eigenvectors, but for each eigenvector it was scaled by the amount of variance in the data explained by that particular eigenvector. This demonstrates that while some authors might receive higher total publication score, their impact on the field might be more concentrated in specific sub-areas.

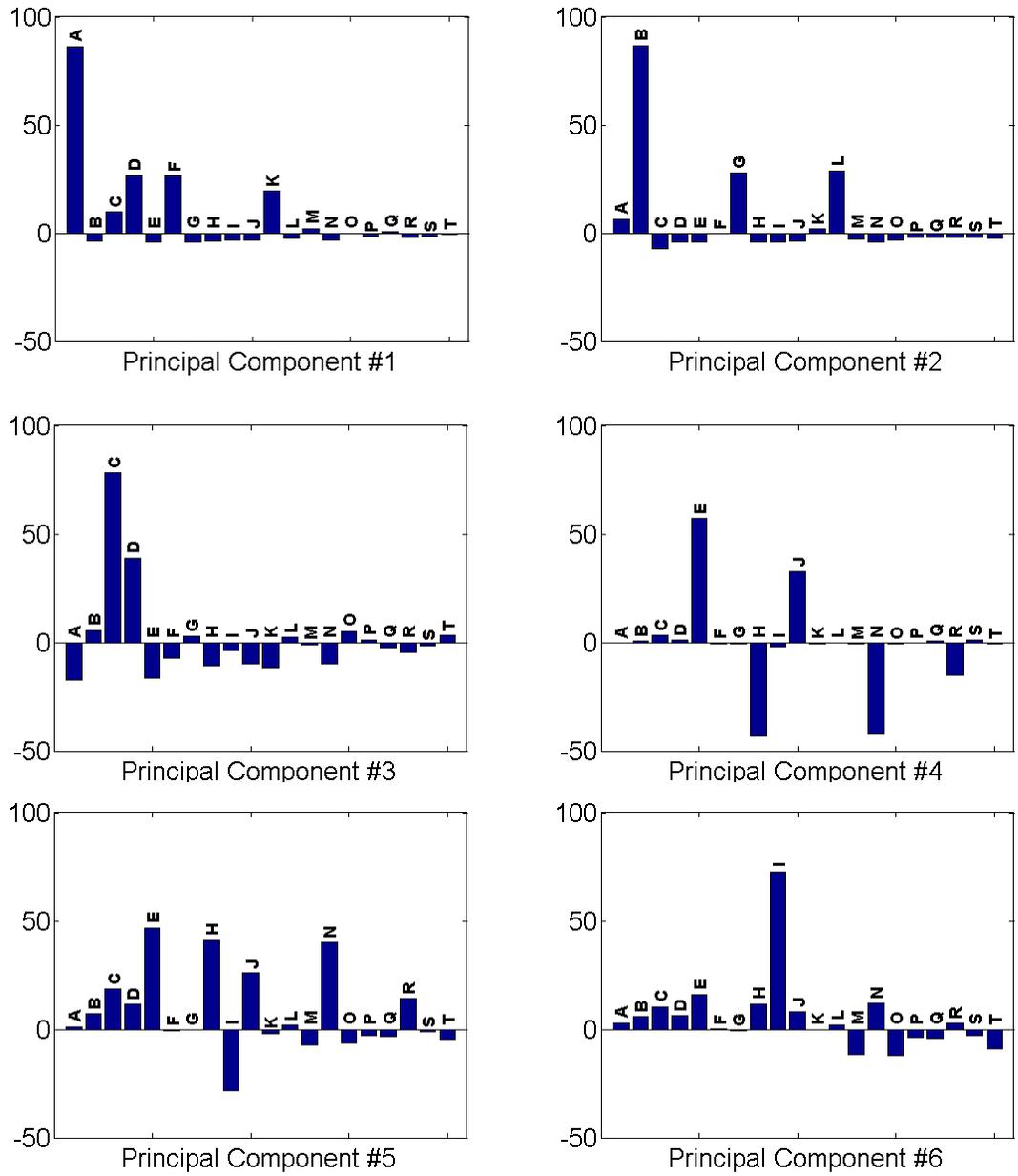

**Figure 1.** Top 20 authors contribution to the first 6 principal components

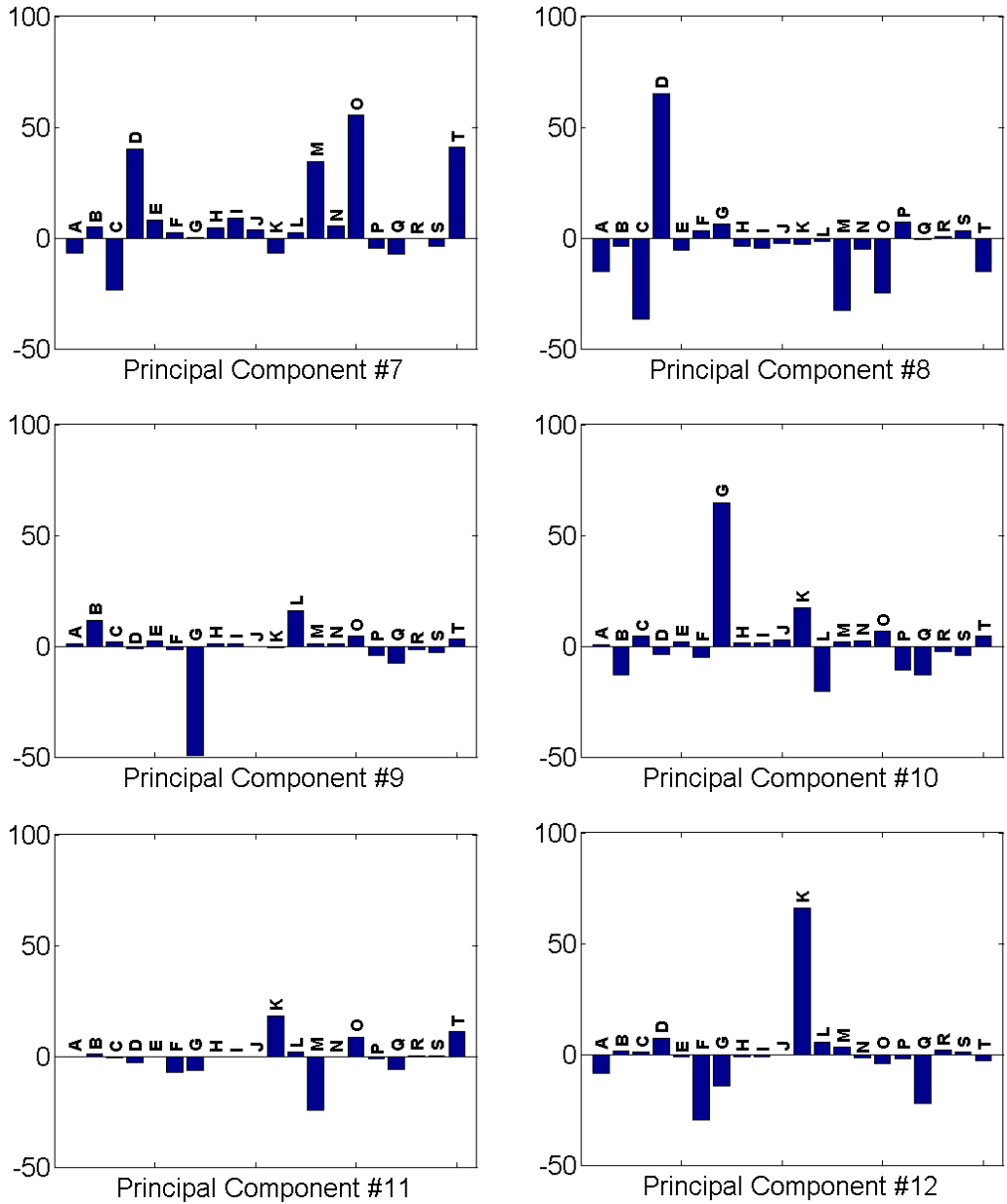

**Figure 2.** Top 20 authors contribution to the second 6 principal components

We subsequently perform the principal component analysis[8, 9] on the full data set. In PCA, the full data set is converted into a linear superposition of orthogonal, linearly uncorrelated eigenvectors $w_k$ in Equation 1 such that

$$P_i(A_j) = a_k w_k(A_j),  \qquad \text{Eq. (1)}$$

Where $a_k$ are expansion coefficients, or component weights and $P_i(A_j)$ is the author list of selected publication. The eigenvectors $w_k(A_j)$ and the corresponding eigenvalues $\lambda_k$ are found from the singular value decomposition of covariance matrix, $\mathbf{C} = \mathbf{A}\mathbf{A}^T$, where $\mathbf{A}$ is the matrix of all publications $\mathbf{A}_{ij}$ i.e. the rows of $\mathbf{A}$ correspond to individual papers ($i = 1,..,N$), and columns correspond to individual authors, $j = 1,..,P$. The eigenvectors $w_k(A_j)$ are orthogonal and are arranged such that corresponding eigenvalues are placed in descending order, $\lambda_1 > \lambda_2 > ....$ by variance. In other words, the first eigenvector $w_1(A_j)$ contains the most information within the spectral image data; the second contains the most common response after the subtraction of variance from the first one, and so on.

The structure of the first 12 eigenvectors is illustrated in Figures 1,2. The simple examination of the associated eigenvectors reveals the structure of interactions and collaborations in the community defined by "piezoresponse". For example, the first eigenvector is dominated by A, and has strong contributions from D, F, and K. This is fully reasonable, since D has been a postdoc with A, F is staff member in his group, and K has been a collaborator for over 10 years. The contribution of C is small but positive, reflecting several years of collaborative effort. Finally, contributions of all other groups are small but negative, as expected for competitive field. The second eigenvector is dominated by B, with strong positive contributions by G and L. Both have been postdoctoral scientists in his group. Similarly, most of the other contributions are weakly negative. The analysis of third eigenvector dominated by C reveals strong contribution by D – again, the latter was his graduate student. Similar analysis can be applied to all other eigenvectors. For example, eigenvector 6 reflects the effort by I group, eigenvector 7 by O group, and 8 independent work by D. The examination of all other eigenvectors similarly delineates work and history of other major piezoresponse groups. Remarkably, this structure can be obtained via simple examination of publication records, without direct knowledge of the field.

The further insight into the structure of community can be derived via clustering of the data. K-means algorithm divides $M$ points in $N$ dimensions into $K$ clusters so that the within cluster the sum of squares, as shown in Equation 2,

$$\arg\min \sum_{i=1}^{k} \sum_{x_j \in S_i} ||x_j - \mu_i||^2, \qquad \text{Eq. (2)}$$

where $\mu_i$ is the mean of points in $S_i$, is minimized.[10, 11] Here, we have used a Matlab k-means algorithm that minimizes the sum, over all clusters, of the within-cluster sums of point-to-cluster-centroid distances. As a measure of distance (minimization parameter) we have used sum of absolute differences with each centroid being the component wise median of the points in a given cluster.

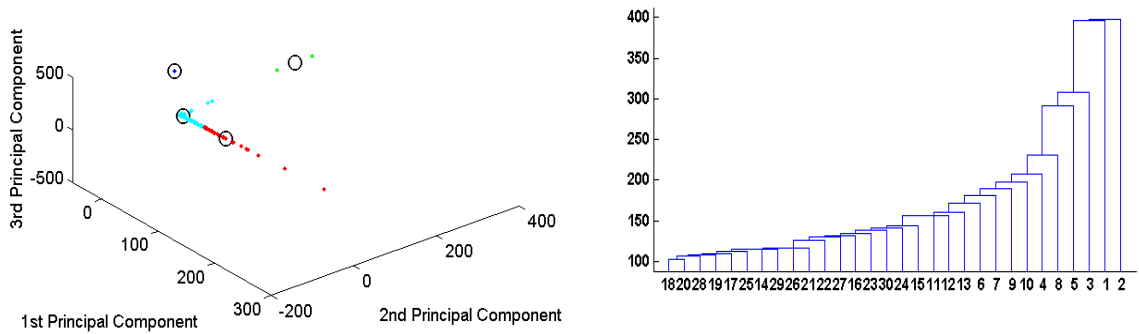

**Figure 3.** Clustering in 2D and 3D space representation of first principal components

Finally, we note that the similar analysis can be used to describe the dynamics of the community, via the integral (i.e. all publications up to certain year) or differential (all publications in a certain year) analysis of the author space. The evolution of the field will then be represented by the dynamics of the eigenvectors with time. We note that differential eigenvectors are often discontinuous (i.e. if relatively publication rate of group changed, the eigenvectors switch). Hence, as one approach to visualize this data, we classify eigenvectors based on strongest elements (aligned with individual groups), and plot the number vs. time. This dynamics is illustrated in Figure 4, and represents the relative changes in the contributions with time.

**Table II**

Time evolution of impact

| Author | Score |
|--------|-------|
| A | 2875.5 |
| B | 1345.75 |
| C | 1197.25 |
| D | 1033 |
| E | 775.25 |
| F | 597.5 |
| G | 549.25 |
| H | 543.5 |

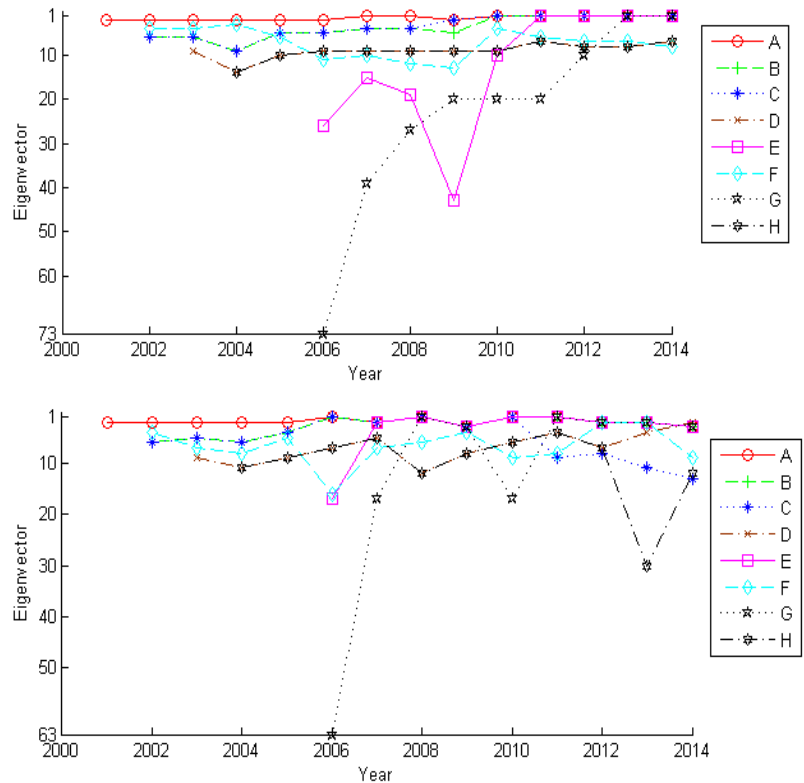

**Figure 2.** Top 8 overall authors with their highest eigenvector result per year. Integral (top) and Differential (bottom)

To summarize, we suggest simple analysis to analyze the structure of scientific community described by a set of keywords. The proposed approach can be used for multiple applications, including choice of optimal reviewers for papers and proposals (different clusters within the same field), describe the rate at which person acquires citations, downloads, get retweeted, or other metric of occupying information space, and define parameters defining overlap between identity of the conference in the field. We note that metrics can be extended to include participation in the conferences or discussions can be parameterized based on time (easiest), contributions (subjective), or results (often difficult to determine and ascribe properly). The advantage of this approach that it relies on simple statistical measures of the publication data, and does not require extensive knowledge of the field available though many year experience.

This approach is also considerably simpler then approaches based on the full semantic analysis of the text documents that potentially allow communities to be defined based on the hierarchical semantic clustering.

Future opportunities include developing algorithms and weighting schemes that favor productive research. It is important to note that this article describes tools for research analysis; however assigning value is more complex. Choice of weighting schemes will have large effect on the strategies scientific community will adopt in publication – e.g. whether the total credit per paper is one or one per author will determine preferred number of authors on the papers. Nevertheless, an integrated policy by scientific governing bodies and funding agencies could then optimize the output of scientific research.

**Acknowledgements:**

Artem Maksov acknowledges fellowship support from the UT/ORNL Bredesen Center for Interdisciplinary Research and Graduate Education. This research was conducted at the Center for Nanophase Materials Sciences, which is a DOE Office of Science User Facility.